\documentclass[superscriptaddress,aps,showpacs,preprintnumbers,nofootinbib,10pt,twocolumn]{revtex4}

\usepackage{graphicx}
\usepackage{amsmath,amssymb}

\begin{document}

\preprint{DF/IST-6.2009, CFTP/09-022}

\title{Constraints on unparticle long range forces from big bang nucleosynthesis bounds on the variation of the gravitational coupling}
\author{O. Bertolami}
\email{orfeu@cosmos.ist.utl.pt} \affiliation{Departamento de F\'{\i}sica and Instituto de Plasmas e Fus\~ao Nuclear, Instituto Superior T\'{e}cnico, Avenida Rovisco
Pais, 1049-001 Lisboa, Portugal}

\author{N. M. C. Santos}
\email{ncsantos@cftp.ist.utl.pt} \affiliation{Centro
de F\'{\i}sica Te\'orica de Part\'{\i}culas, Instituto Superior T\'{e}cnico, Avenida Rovisco
Pais, 1049-001 Lisboa, Portugal}

\begin{abstract}

We use big bang nucleosynthesis bounds on the variation of the gravitational coupling to derive constraints on the strength of the deviation from the gravitational inverse-square law due to tensor and vector unparticle exchange.

\end{abstract}

\date{\today}


\maketitle

\centerline{{\bf Dedicated to the memory of M. C. Bento}}

\section{Introduction}

Recently, the possibility of the existence of new physics above the TeV scale has been considered through the introduction of unparticles~\cite{Georgi:2007ek,Georgi:2007si}. In this scheme one admits a hidden sector with a nontrivial infrared fixed point $\Lambda_U$, below which scale invariance is explicit. In the ultraviolet (UV) regime, at energies above $\Lambda_U$, the hidden sector operator ${\cal O}_{UV}$ of dimension $d_{UV}$ couples to the standard model (SM) fields through an operator ${\cal O}_{SM}$ of dimension $n$ via nonrenormalizable interactions ${\cal O}_{UV}\,{\cal O}_{SM}/ M_U^{d_{UV}+n-4}$, where $M_U$ is the mass of the heavy exchanged particle. Below $\Lambda_U$, the hidden sector becomes scale invariant and the operator ${\cal O}_{UV}$ mutates into an unparticle operator ${\cal O}_{U}$ with noninteger scaling dimension $d_u$. The coupling of field operators can be generically written as
\begin{align}
\dfrac{\Lambda_U^{d_{UV}-d_u}}{M_U^{d_{UV}+n-4}}{\cal O}_{U}\,{\cal O}_{SM}~.
\end{align}

The operator ${\cal O}_U$ could be a scalar, a vector, a tensor or even a spinor. Collider signatures~\cite{Cheung:2007zza,Cheung:2007ap} as well as other phenomenological aspects~\cite{Luo:2007bq,Chen:2007vv,Ding:2007bm,Liao:2007bx,Aliev:2007qw,Li:2007by,Lu:2007mx,Fox:2007sy,Greiner:2007hr,Chen:2007qr,Kikuchi:2007qd,Lenz:2007nj,Delgado:2007dx,Anchordoqui:2007dp} resulting from this scenario have been investigated. Also, several cosmological~\cite{Davoudiasl:2007jr,McDonald:2007bt,Kikuchi:2007az,Grzadkowski:2008xi} and astrophysical~\cite{Hannestad:2007ys,Das:2007nu,Freitas:2007ip} constraints on unparticle physics have been studied, including bounds arising from stellar equilibrium~\cite{Bertolami:2009aa} and black hole evolution~\cite{Dai:2008qn}. The exchange of unparticles gives rise to long range forces which deviate from the inverse-square law (ISL) for massless particles due to the anomalous scaling of the unparticle propagator. For example, the exchange of scalar (pseudoscalar) unparticles can give rise to spin-dependent long range forces, as pointed out in Ref.~\cite{Liao:2007ic}. Coupling between unparticles and vector or axial-vector currents have been investigated in Ref.~\cite{Deshpande:2007mf}. In~Ref.~\cite{Goldberg:2007tt} the coupling between unparticles and the energy-momentum tensor was studied . Torsion-balance experiment results have been used to put limits in these interactions~\cite{Freitas:2007ip,Deshpande:2007mf,Barranco:2009px,Goldberg:2007tt}. In this work we will examine the deviations from ISL due to tensor and vector particle exchange and the possible constraints that can be derived using the bounds on the variation of the gravitational coupling $G$, at the time of big bang nucleosynthesis (BBN).

\section{Ungravity from tensor unparticles}

If ${\cal O}_U$ is a rank-two tensor it could couple to the stress-energy tensor $T_{\mu \nu}$ and its exchange between physical particles could lead to a modification of Newtonian gravity. Taking the gravitational coupling of the tensor unparticle to $T_{\mu \nu}$ to be of the form
\begin{align}
\dfrac{1}{M_\star\,\Lambda_U^{d_u-1}}\, \sqrt{g}\; T_{\mu \nu} \,{\cal O}_U^{\mu \nu}~,
\end{align}
where $M_\star=\Lambda_U \left(M_U/\Lambda_U \right)^{d_{UV}}$, it can be shown that, in the nonrelativistic limit and for $d_u \neq 1$, the effective gravitational potential of the unparticle exchange has the form~\cite{Goldberg:2007tt}
\begin{align}
V(r)=-G_N \,\dfrac{m_1\, m_2}{r}\left[1+\left(\dfrac{R_G}{r}\right)^{2 d_u-2}\right]~.\label{potential}
\end{align}
Here $G_N$ should be identified with the Newtonian gravitational constant, $G_N=6.7 \times 10^{-39}$~GeV$^{-2}$. The characteristic length scale $R_G$ for which the ungravity interactions become significant is defined to be
\begin{align}
R_G=\dfrac{1}{\Lambda_U}\,\left(\dfrac{M_{Pl}}{M_\star}\right)^{\frac{1}{d_u-1}}\, C(d_u)^{\frac{1}{2d_u-2}}~,\label{RG}
\end{align}
where $M_{Pl}=1.22 \times 10^{19}$GeV is the Planck mass, and $C(d_u)$ is given by
\begin{align}
C(d_u)= \dfrac{2}{\pi^{2d_u-1} }\dfrac{\Gamma(d_u+\frac{1}{2})\Gamma(d_u-\frac{1}{2})}{\Gamma(2d_u)}~.\label{Calpha}
\end{align}

The case $d_u<1$ leads to forces which fall slower than gravity and can be easily ruled out from fifth force experiments~\cite{Adelberger:2006dh}. Hence we will consider only $d_u>1$ (see, however, Ref.~\cite{Bertolami:2009aa}). Torsion-balance experiment searches for power-law modifications to the ISL have been used to constrain the modified potential~\cite{Goldberg:2007tt}. Also, the observed perihelion precession of Mercury has been used to test these interactions~\cite{Das:2007cc}.

\begin{figure}[t]
  \includegraphics[width=8.0cm]{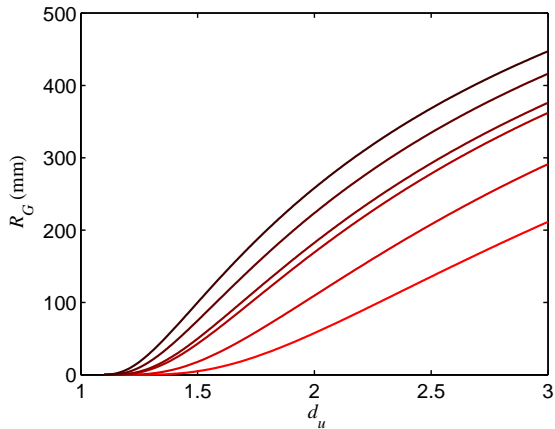}\\
  \caption{(color online) Upper bound on $R_G$ as function of $d_u$ for  $\left|\Delta G/G\right| \leq 0.2,~0.15,~0.1,~0.086,~0.036,~0.01$ (from top to bottom).}
  \label{fig1}
\end{figure}

\begin{figure*}[t]
  \includegraphics[width=8.0cm]{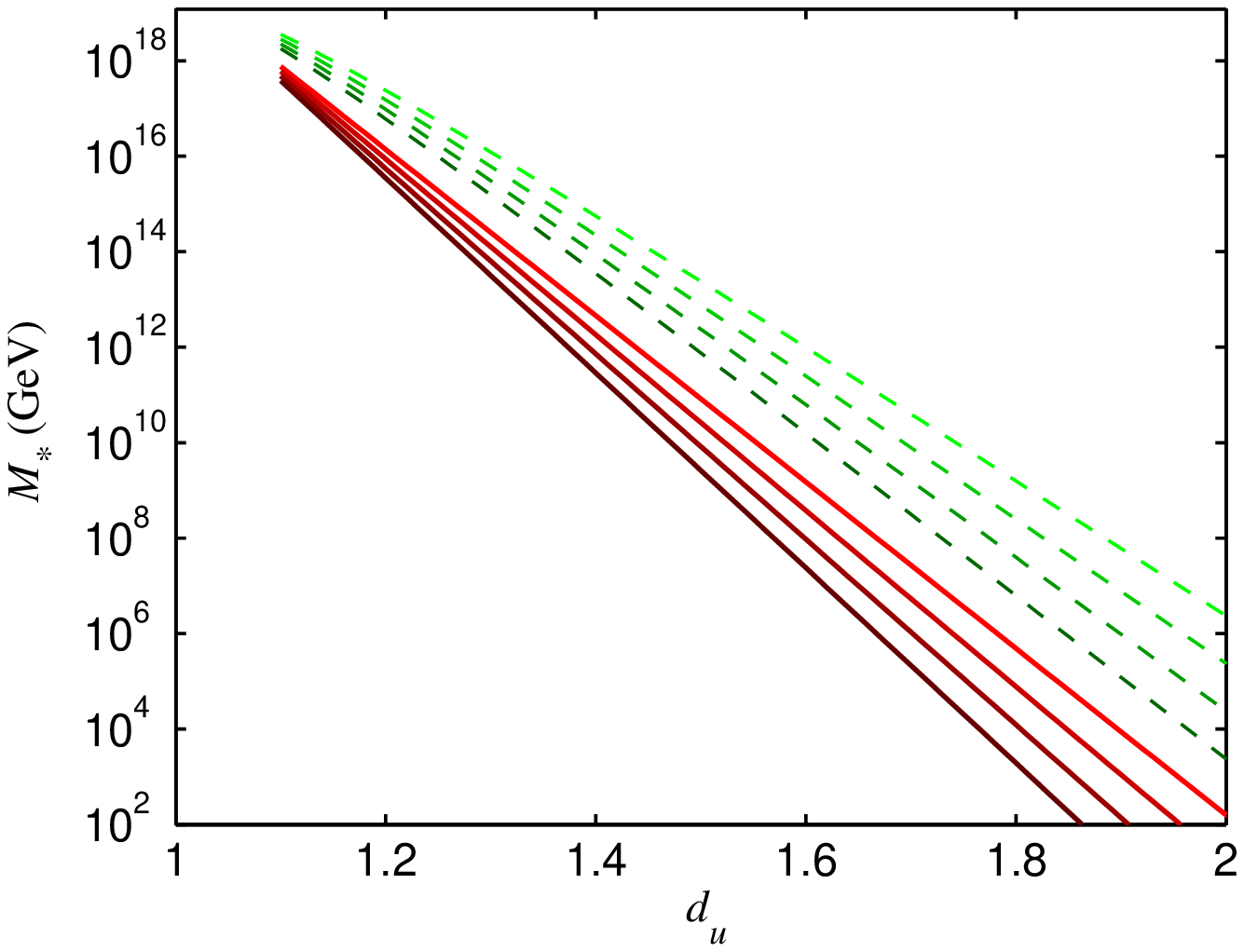}
  \includegraphics[width=8.0cm]{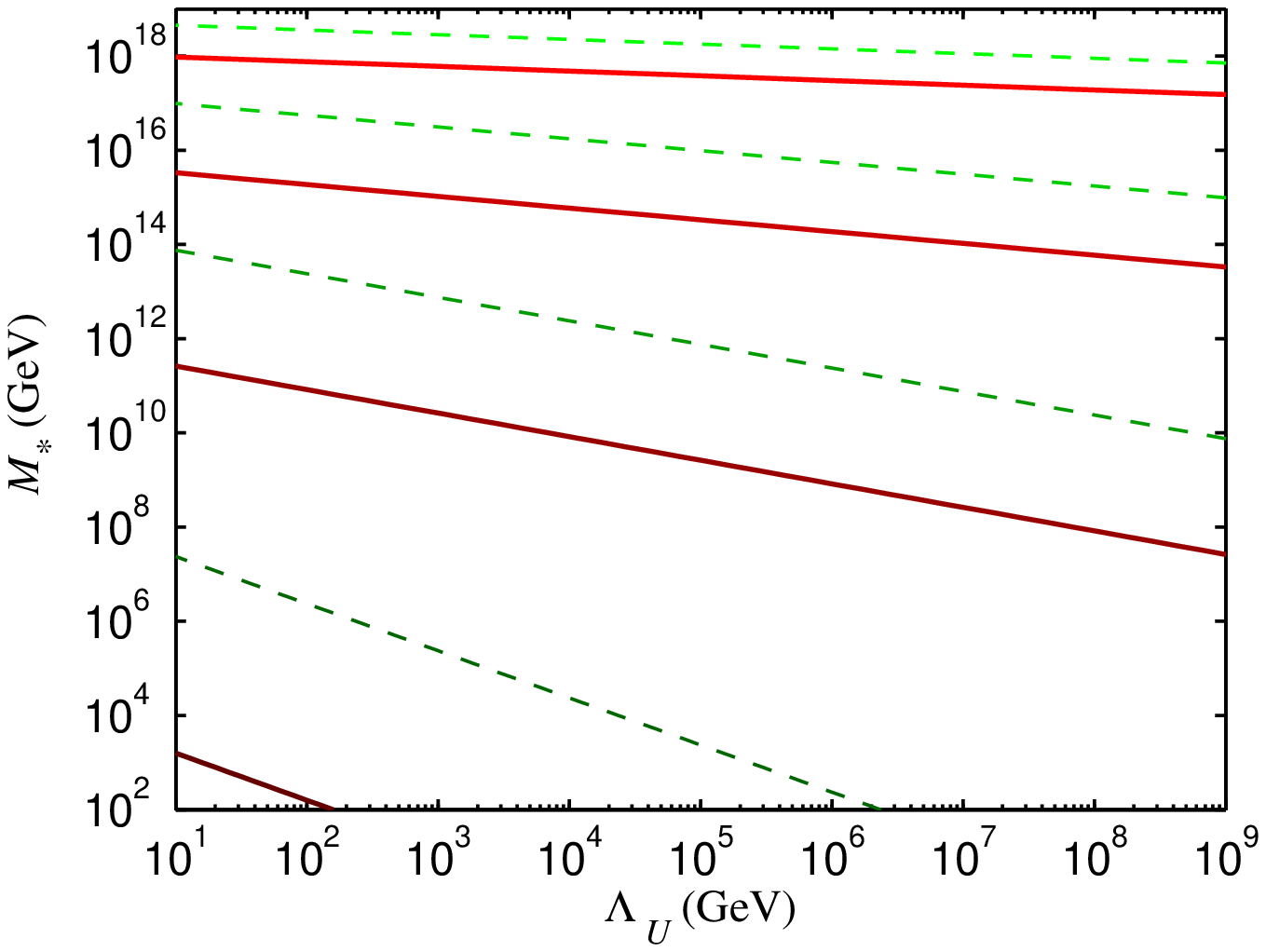}\\
  \caption{(color online) Allowed region (above the curves) on the $M_\star - d_u$ (left) and $M_\star-\Lambda_u$ (right) parameter spaces for $\Delta G/G \leq 0.086$. On the left it has been taken $\Lambda_U= 100$~GeV, $1,~10,~100$~TeV [lighter (top) to darker (bottom) curves] and on the right $d_u=1.1,~1.25,~1.5$~and $2$ [lighter (top) to darker (bottom) curves] obtained from BBN bounds (red solid lines) and ISL violation data (green dashed lines).}
  \label{fig2}
\end{figure*}

The modification of the gravitational potential can be seen as a dependence of the gravitational coupling on $r$. Indeed, the force associated with the potential of Eq.~(\ref{potential}) is
\begin{align}
\mathbf{F}(r)=-\mathbf{\nabla} V(r) = -G(r)\, \dfrac{m_1\, m_2}{r^2}\, \hat{\mathbf{r}}~,
\end{align}
with an effective gravitational coupling given by
\begin{align}
G(r)=G_N\left[1+(2d_u-1)\left(\dfrac{R_G}{r}\right)^{2 d_u-2}\right]~.
\end{align}

In this work we investigate the possible limits on the different energy scales ($\Lambda_U$ and $M_\star$) that can be derived using the bounds on the variation of the gravitational coupling $G$,
\begin{align}
\left|\dfrac{\Delta G}{G}\right| \equiv \left|\dfrac{G(r)-G_N}{G_N}\right|=(2d_u-1)\left(\dfrac{R_G}{r}\right)^{2 d_u-2}~,\label{G of r}
\end{align}
at the time of BBN. The effect of a varying $G$ on BBN is through the Hubble expansion law, $H \equiv \dot{a}/a \propto \sqrt{G}$ ($a$ being the scale factor), which determines both the neutron-proton density ratio at freeze-out and the efficiency of $^2$H burning abundance when nucleosynthesis starts. Given the large statistical and systematic errors of the measurements, the typical constraints on the variation of $G$ are of the order of a few percent. In particular, it is found that~\cite{Iocco:2008va}
\begin{align}
-0.036 \leq \dfrac{\Delta G}{G} \leq 0.086~,
\end{align}
at $95\%$ confidence level (C.L.), where the values $Y_p = 0.250 \pm 0.003$ and $^2$H/H$=2.87^{+0.22}_{-0.021}$ have been used for the $^4$He mass fraction and the normalized deuterium number density, respectively. The result is dominated by the effect on $^4$He. Similar results are found by several other authors, depending on the experimental values for light nuclei adopted in their analysis (see~\cite{Iocco:2008va} and references therein). In the pessimistic case of a large systematic error on $Y_p$, deuterium can provide a $20\%$ bound~\cite{Iocco:2008va}.

In order to infer on the range and strength of the unparticle force one needs to know the typical distance $r$ between particles interacting during BBN. It turns out that during this epoch neutrons and protons have mean free paths of the same order of magnitude, $\lambda_p \sim \lambda_n \sim \lambda= 1\,$m, which in the following we take as the typical distance $r$ between nucleons~\cite{Bertolami:1999sj,Applegate:1987hm}.

From Eq.~(\ref{G of r}) it is now easy to derive upper bounds on the characteristic length scale $R_G$ for which the ungravity interactions become significant. On Fig.~\ref{fig1} the upper bounds on $R_G$ as a function of $d_u$ are depicted, considering $\left|\Delta G/G\right| \leq 0.2,~0.15,~0.1,~0.086,~0.036$ and $0.01$ . On Fig.~\ref{fig2} we show the allowed regions (above the solid lines) on the $M_\star - d_u$ (left panel) and $M_\star - \Lambda_U$ (right panel) parameter spaces, which can be derived from Eq.~(\ref{RG}) when combined with the previous constraint. On the left panel we present lines for $\Lambda_U= 100$~GeV, $1,~10,~100$~TeV; and on the right one it has been considered $d_u=1.1,~1.25,~1.5$~and $2$. In both cases we have fixed $\Delta G/G \leq 0.086$.

These limits should be compared to the ones that can be derived from ISL deviations using precision submillimeter tests. Following~\cite{Goldberg:2007tt}, we use the data from Ref.~\cite{Adelberger:2006dh}, where the potential
\begin{align}
V_{12}(r)=-G_N \,\dfrac{m_1\, m_2}{r}\left[1+\beta_k\left(\dfrac{1 {\rm mm}}{r}\right)^{k-1}\right]~
\end{align}
is considered, to obtain an upper limit on $R_G$, by identifying $d_u=(k+1)/2$ and $R_G=\beta_k^{1/(k-1)}$. From that upper bound we easily get the constraints on the energy scales. Those limits are plotted on Fig.~\ref{fig2} (dashed lines), together with the ones obtained from BBN. We interpolated the limits in Table I of Ref.~\cite{Adelberger:2006dh} to obtain bounds on $R_G$ as a function of $d_u$. For $k \leq 3$ we consider $\beta_k$ as a function of $1/(k-1)^2$, while for larger $k$ we simply interpolated the limit linearly.

We find that these bounds on $M_\star$ are stronger than the ones obtained from BBN. Notice however that, unlike the laboratory bounds, our results test ungravity at the early universe.

\section{Long range forces due to vector unparticles}

Let us now consider the long range forces resulting from the coupling of vector unparticles~\cite{Deshpande:2007mf}. The potential for a coupling between a vector unparticle and a baryonic (or leptonic) current $J_\mu$ of the form
\begin{align}
\dfrac{\lambda}{\Lambda_U^{d_u-1}}\, J_{\mu} \,{\cal O}_U^{\mu}~,
\end{align}
is given by~\cite{Deshpande:2007mf}
\begin{align}
V_U= \dfrac{\lambda^2 \,N_1\, N_2\,  \tilde{C}\, (d_u)}{\Lambda_U^{2d_u-2}}\dfrac{1}{r^{2 d_u-1}}~,
\end{align}
where $N_{1,2}$ are the total number of baryons of the two interacting objects and
\begin{align}
 \tilde{C} (d_u)= \dfrac{1}{2 \pi^{2d_u} }\dfrac{\Gamma(d_u+\frac{1}{2})\Gamma(d_u-\frac{1}{2})}{\Gamma(2d_u)}~.\label{Cprime}
\end{align}
Combined with the gravitational potential we can write
\begin{align}
V(r)=-G_N \,\dfrac{m_1\, m_2}{r}\left[1-\left(\dfrac{\tilde{R}_G}{r}\right)^{2 d_u-2}\right]~,\label{potentialvector}
\end{align}
with
\begin{align}
\tilde{R}_G= \dfrac{1}{\Lambda_U}\,\left(\dfrac{\lambda\,M_{Pl}}{u}\right)^{\frac{1}{d_u-1}}  \tilde{C} (d_u)^{\frac{1}{2d_u-2}}~.
\end{align}
where, in order to obtain numerical results, we have made the approximation $N_{1,2} \approx m_{1,2}/u$, $u=931.4$~MeV being the atomic mass unit.

\begin{figure*}[tbh]
  \includegraphics[width=8.0cm]{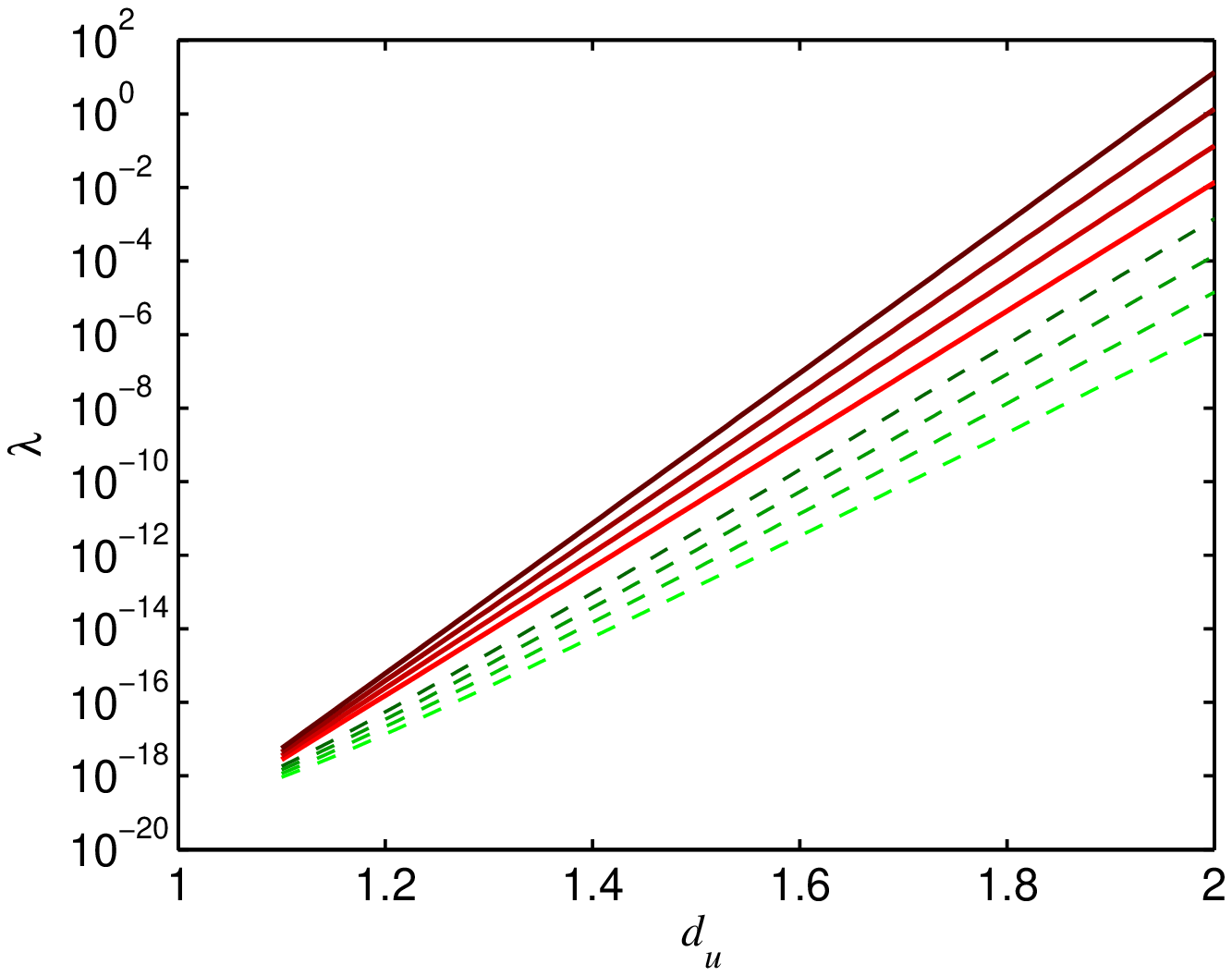}
  \includegraphics[width=8.0cm]{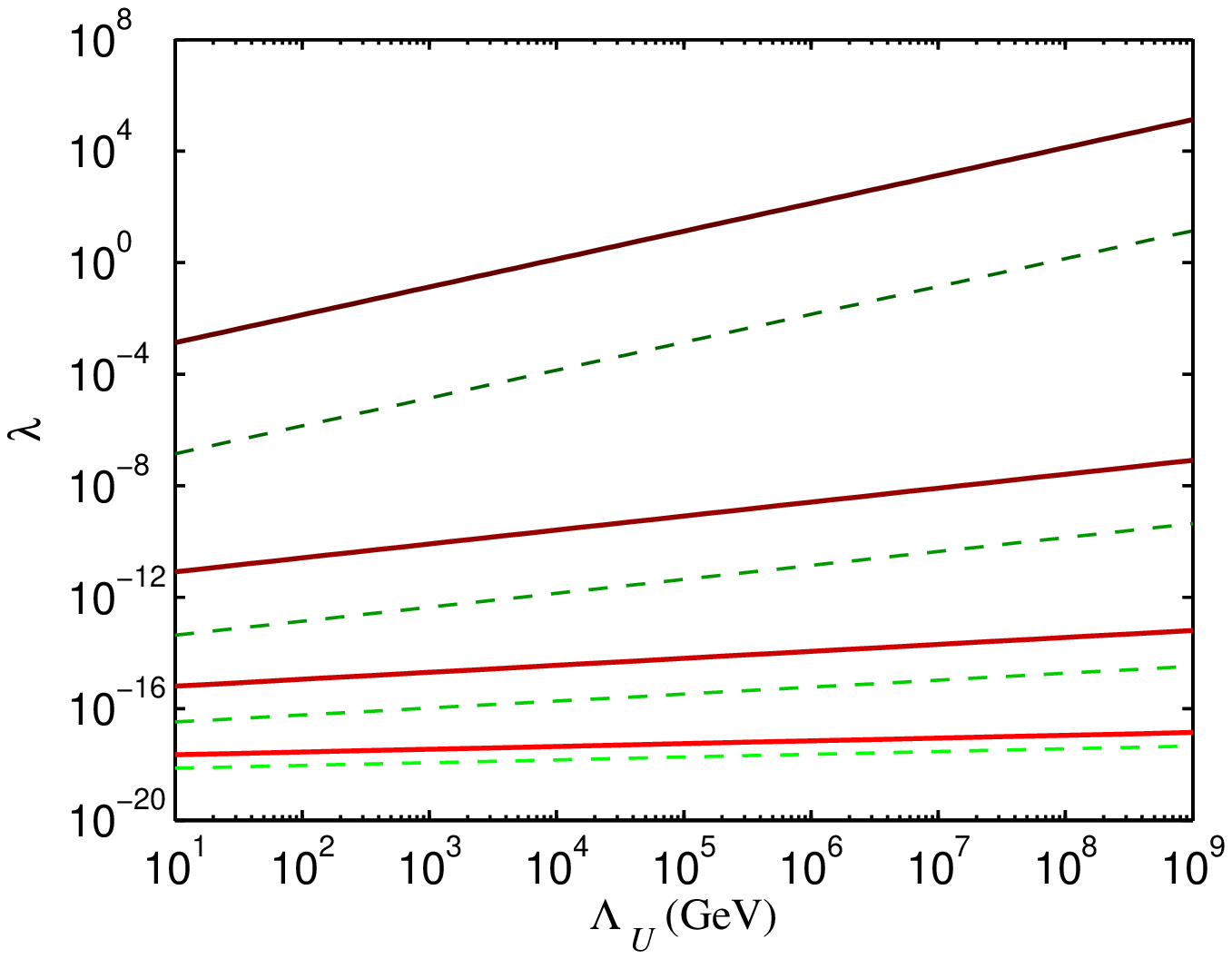}\\
  \caption{(color online) Allowed region (below the curves) on the $\lambda - d_u$ (left) and $\lambda-\Lambda_u$ (right) parameter spaces for $\Delta G/G \geq -0.036$. On the left we took $\Lambda_U= 100$~GeV, $1,~10,~100$~TeV [lighter (bottom) to darker (top) curves] and on the right $d_u=1.1,~1.25,~1.5$~and $2$ [lighter (bottom) to darker (top) curves] obtained from BBN bounds (red solid lines) and ISL violation data (green dashed lines).}
  \label{fig3}
\end{figure*}

The vector unparticle exchange causes a negative variation of $G$, given that the force is repulsive. Hence in this case we consider the bound $\Delta G/G \geq -0.036$. Following the line of thought of the previous section we are able to derive the limits on $\lambda$ presented in Fig.~\ref{fig3}, where the solid (dashed) lines represent the upper bounds on this coupling as a function of $d_u$ and $\Lambda_U$ resulting from BBN (torsion-balance experiments) constraints. As in the tensor exchange case, we find that the bounds arising from laboratory searches of putative violations of the ISL are more stringent.

\section{Conclusions}

In this work we have examined the importance of the existence of unparticles on BBN yields through the modification they introduce in the ISL. We have considered tensor and vector unparticle exchange.

We find that in both cases the BBN bounds are less stringent than the laboratory ones searching for violations of the ISL. For $d_u$ close to unity, the bounds are comparable. From the constraints on the variation of $G$ during BBN, considering $\Lambda_U =1$~TeV and $d_u=1.1$, we find $M_\star \geq 6.04 \times 10^{17}$~GeV and $\lambda \leq 3.54 \times 10^{-18}$, for tensor and vector exchange, respectively, while fifth force experiments yield $M_\star \geq 2.83 \times 10^{18}$~GeV and $\lambda \leq 1.17 \times 10^{-18}$. We should notice that the range $\Lambda_U \gtrsim 1$~TeV is the most interesting case to consider since it is the one where unparticles may be observed in future colliders.

The difference between BBN and laboratory bounds becomes more visible for larger values of $d_u$. For $d_u=2$, we get $M_\star \geq 15.9$~GeV and $\lambda \leq 1.34 \times 10^{-1}$ from BBN, and  $M_\star \geq 2.36 \times 10^{5}$~GeV and $\lambda \leq 1.40 \times 10^{-5}$ from torsion-balance experiments. We remark however that our bounds concern unparticle effects on the early universe and, given the methodological differences, should be regarded as complementary to the laboratory ones.

\vspace*{4mm}

\textbf{Acknowledgments}\\
N.M.C.S. acknowledges support of the Funda\c{c}\~{a}o
para a Ci\^{e}ncia e a Tecnologia (FCT, Portugal) through Grant No. SFRH/BPD/36303/2007.

\end{document}